\documentclass[aip,reprint,amsmath,amssymb,natbib]{revtex4-1}

\usepackage{graphicx,color}
\usepackage{dcolumn}
\usepackage{bm}
\renewcommand{\vec}[1]{\mathbf{#1}}

\begin{document}

\preprint{AIP/123-QED}

\title[]{Numerical simulations of strong incompressible magnetohydrodynamic turbulence}

\author{J. Mason}\affiliation{Department of Astronomy \& Astrophysics, University of Chicago, 5640 S. Ellis Ave, Chicago, IL, 60637, USA}
\author{J.C. Perez}\affiliation{Institute for the Study of Earth, Oceans, and Space, University of New Hampshire, Morse Hall, 8 College Road, Durham, NH, 03824}
\author{S. Boldyrev}\affiliation{Department of Physics, University of Wisconsin at Madison, 1150 University Ave, Madison, WI 53706, USA}
\author{F. Cattaneo}\affiliation{Department of Astronomy \& Astrophysics, University of Chicago, 5640 S. Ellis Ave, Chicago, IL, 60637, USA}

\date{\today}

\begin{abstract}
Magnetised plasma turbulence pervades the universe and is likely to play an important role in a variety of astrophysical settings. Magnetohydrodynamics (MHD) provides the simplest theoretical framework in which phenomenological models for the turbulent dynamics can be built. Numerical simulations of MHD turbulence are widely used to guide and test the theoretical predictions; however, simulating MHD turbulence and accurately measuring its scaling properties is far from straightforward. Computational power limits the calculations to moderate Reynolds numbers and often simplifying assumptions are made in order that a wider range of scales can be accessed. After describing the theoretical predictions and the numerical approaches that are often employed in studying strong incompressible MHD turbulence, we present the findings of a series of high-resolution direct numerical simulations. We discuss the effects that insufficiencies in the computational approach can have on the solution and its physical interpretation.
\end{abstract}

\maketitle

\section{\label{sec:Introduction} Introduction}

Magnetohydrodynamic turbulence constitutes one of the most important unresolved problems in classical physics.
Magnetohydrodynamics provides the simplest theoretical framework in which we can develop our understanding of magnetised plasma turbulence, which itself forms the foundation for a vast array of astrophysical phenomena that are believed to be magnetically driven. However, despite it being almost 50 years since the pioneering work of Iroshnikov~\cite{iroshnikov_64} \& Kraichnan~\cite{kraichnan_65}, debates continue over even the most fundamental issues, such as the inertial range scaling properties of the energy cascade. 

We consider here the simplest case of incompressible field-guided MHD turbulence. It is useful to write the equations governing the evolution of the fluctuating velocity $\vec{v}(\vec{x},t)$ and magnetic field $\vec{b}(\vec{x},t)$ in terms of the Els\"asser variables $\vec z^\pm=\vec v\pm\vec b$,
\begin{equation}
 \label{eq:mhd-elsasser}
  \left( \frac{\partial}{\partial t}\mp\vec V_A\cdot\nabla \right) \vec
  z^\pm+\left(\vec z^\mp\cdot\nabla\right)\vec z^\pm = -\nabla P +\nu\nabla^2 \vec z^{\pm}+\vec f^\pm, 
  \end{equation}
  \begin{equation}
  \label{eq:mhd-div}
  \nabla \cdot {\vec z}^{\pm}=0 
  \end{equation}
where $\vec{V_A}=\vec{B_0}/\sqrt{4\pi\rho_0}$ is the Alfv\'en velocity based on the uniform background magnetic field $\vec{B_0}$, $\rho_0$ is the background plasma density that we assume to be constant, $P=(p/\rho_0+b^2/2)$ is the total pressure, $\nu$ is the fluid
viscosity (which for simplicity has been taken to be equal to the magnetic diffusivity) and $\vec f^\pm$ represents random forces that drive the
turbulence at large scales. The pressure term can be eliminated in favour of a projection of the solution onto the `incompressible plane'  (see, e.g., Ref.~\onlinecite{lesieur_97}). 
 In the absence of forcing and dissipation the linearised system admits solutions in the form of Alfv\'en waves that travel parallel or antiparallel to the background field $\vec{B_0} =B_0 \vec{\hat e_z}$, say, with the Alfv\'en speed. A normal mode analysis reveals the  dispersion relation $\omega^\pm(\vec k)=\pm k_\| V_A$. The waves are transverse and they can be divided into two classes: shear Alfv\'en waves with
polarizations perpendicular to both $\vec{B_0}$ and to the wavevector
$\vec{k}$, and pseudo Alfv\'en waves with polarizations in the plane of $\vec{B_0}$ and $\vec{k}$, perpendicular to $\vec{k}$.

Interactions between counter-propagating
Alfv\'en wave packets transfer energy to smaller scales (Ref.~\onlinecite{kraichnan_65})
until eventually the dissipative scales are reached and energy is 
removed from the system. The efficiency of the
transfer is controlled by the relative size of the linear
and nonlinear terms in equation (\ref{eq:mhd-elsasser}). The regime in
which the linear terms dominate is known as
{\it weak} MHD turbulence, 
otherwise the turbulence is called {\it
  strong}. In fact, it has been demonstrated both analytically and
numerically that the energy cascade occurs predominantly in the plane
perpendicular to the guiding magnetic field
(Refs.~\onlinecite{shebalin83,maron_g01}), which ensures that even if the turbulence
is weak at large scales it encounters the strong regime as the cascade
proceeds to smaller scales. We concentrate here on the strong case.

The first phenomenological theories for the energy cascade in strong incompressible
MHD turbulence were devoted to the so-called balanced case, in which the energies
in each of the Els\"asser fields $E^\pm=\frac{1}{4}\int
(z^\pm)^2 d^3 x$ are approximately equal and hence there is no net cross helicity 
in the system, $H^c=\int ({\bf v}\cdot {\bf b}) d^3 x= E^+ - E^- \approx 0$. 
For this case, Goldreich \& Sridhar~\cite{goldreich_s95} argued that as the cascade proceeds to smaller
scales the linear and nonlinear terms establish a so-called `critical balance'
\begin{equation}
\label{eq:critical_balance}
V_Ak_\| \sim z^{\mp}_{\perp} k_{\perp}.
\end{equation}
Consequently, the nonlinear interaction time $\tau \sim \lambda_{\perp}/z^\mp_{\perp}$ balances the Alfv\'en time $\tau_A \sim \lambda_{\|}/V_A$ and it follows that the total field-perpendicular energy spectrum $E(k_{\perp}) \propto k_{\perp}^{-5/3}$, with the field-parallel and field-perpendicular lengthscales related by $\lambda_{\|} \propto \lambda_{\perp}^{2/3}$ (Ref.~\onlinecite{goldreich_s95}). Although numerical simulations at the time verified the anisotropic cascade (e.g.~Refs.~\onlinecite{cho_v00, milano_mdm01, maron_g01,muller_bg03}),  doubts began to surface in later years when higher resolution simulations with stronger guide fields seemed to be more consistent with $E(k_{\perp}) \propto k_{\perp}^{-3/2}$ (e.g.~Ref.~\onlinecite{maron_g01, muller_bg03,muller_g05,haugen_bd04a, mason_cb08}). An explanation was provided by the theory of scale-dependent dynamic alignment~\cite{boldyrev_05,boldyrev_06}. The theory predicts that the fluctuating velocity and magnetic fields align within a small angle $\theta \propto \lambda^{1/4}$ in the plane perpendicular to the background field. Alignment reduces the size of the nonlinear term by an amount proportional to $\theta$ and yields  $E(k_{\perp}) \propto k_{\perp}^{-3/2}$. Numerical simulations of driven, globally balanced, field-guided MHD turbulence have verified the alignment scaling~\cite{mason_cb06,mason_cb08,mason_etal11} and have shown that the domain fragments into regions of highly aligned and anti-aligned velocity and magnetic field fluctuations~\cite{boldyrev_mc09,perez_b09}. Thus, even in the globally balanced case, cross-helicity plays a crucial role locally.

Recently, attention has moved on to the globally unbalanced (or cross-helical) system in which $E^+ \ne E^-$. In this case the timescales for the nonlinear deformation of the $z^\pm$ wavepackets  $\tau^\pm \sim \lambda_{\perp}/z_{\perp}^\mp$ can be considerably different. There are a number of competing theoretical predictions that differ in regards to what is assumed about the physics of the nonlinear cascade. For example, the two different theories by Lithwick et al.~\cite{lithwick_gs07} and Beresnyak \& Lazarian~\cite{beresnyak_l08} ultimately arrive at the same conclusion that in the unbalanced regions the field-perpendicular Els\"asser spectra have the same scalings $E^\pm(k_{\perp}) \propto k_{\perp}^{-5/3}$. The two different theories by Perez \& Boldyrev~\cite{perez_b09} and Podesta \& Bhattacharjee~\cite{podesta_bh10} that are based on scale-dependent alignment propose that $E^\pm(k_{\perp}) \propto k_{\perp}^{-3/2}$, while the theory by Chandran~\cite{chandran_08a} concludes that the two Els\"asser spectra have different scalings depending on the degree of imbalance. 

The goal of numerical simulations is to help clarify the picture. The aim is to measure the inertial range scaling properties and to use the results to discriminate between the conflicting theoretical predictions. However, although the task appears to be straightforward, the inference of scaling laws from numerical data is fraught with problems. The two main difficulties result from the difference between the theoretical predictions for the spectral exponents being very small, and from the fact that even state-of-the-art technological resources still limit the extent of the inertial range to approximately a decade. It is therefore of paramount importance to be able to make accurate numerical measurements, to ensure that any source of error in the numerical data is minimised, and to invest all of the computational power in reaching an inertial range that is as extended as possible. In the next section we describe a number of aspects of the simulation design and the techniques for data analysis that enable us to measure the inertial range characteristics as accurately as possible with the currently available computational power.

\section{\label{sec:Formulation} Computational Approach}

\subsection{Simplified equations}

Considerable simplifications can be made by making use of the structure of field-guided MHD turbulence. Since strong MHD turbulence is dominated by fluctuations with $k_{\perp} \gg k_{\|}$, the polarization of the pseudo Alfv\'en waves are closely aligned with $\vec{B_0}$. Since field-parallel gradients are small, Goldreich \& Sridhar~\cite{goldreich_s95} argued that the pseudo Alfv\'en modes are likely to be dynamically insignificant. Indeed, if in equation~(\ref{eq:mhd-elsasser}) we neglect the term $(\vec{z_{\|}} \cdot \nabla_{\|}) \vec{z_{\perp}}$ in comparison with  $(\vec{z_{\perp}} \cdot \nabla_{\perp}) \vec{z_{\perp}}$ then the equations for the shear Alfv\'en dynamics decouple from the pseudo Alfv\'en  dynamics and we obtain
\begin{equation}
 \label{eq:rmhd-elsasser}
  \left( \frac{\partial}{\partial t}\mp\vec V_A\cdot\nabla_{\|} \right) \vec
  z_{\perp}^\pm+\left(\vec z_{\perp}^\mp\cdot\nabla_{\perp}\right)\vec z_{\perp}^\pm = -\nabla_{\perp} P +\nu\nabla^2 \vec z_{\perp}^{\pm}+\vec f_{\perp}^\pm, 
\end{equation}
\begin{equation}
\label{eq:rmhd-div}  
    \nabla \cdot \vec z_{\perp}^{\pm}=0.
  \end{equation}
Equations~(\ref{eq:rmhd-elsasser},\ref{eq:rmhd-div}) are equivalent to the reduced MHD (RMHD) model that was originally derived in the context of fusion devices by Refs.~\onlinecite{kadomtsev_p74, strauss_76} (see also Ref.~\onlinecite{biskamp_03}).
We note that while this system has only two vector components $\vec{z^{\pm}_{\perp}}=(z^\pm_x,z^\pm_y,0)$ each component is a function of all three spatial coordinates, $x,y$ and $z$. Indeed, as stated in the critical balance condition~(\ref{eq:critical_balance}), the linear term involving field-parallel gradients balances the nonlinear term involving field-perpendicular gradients. The RMHD system is therefore fundamentally different than the two-dimensional system in which $\partial_z \equiv 0$.

The reduced MHD model has been widely used in the recent literature for studying the characteristics of strong field-guided MHD turbulence (see, e.g., Refs.~\onlinecite{gomez_md05, perez_b10_2,beresnyak_11}). Compared with system~(\ref{eq:mhd-elsasser}), computing the Fourier series solution to system (\ref{eq:rmhd-elsasser}) is approximately twice as computationally efficient. Such savings can then be invested in reaching the higher Reynolds number regime. In \S\ref{sec:compare} below we directly compare the solutions of the RMHD model and the full MHD equations. We verify and put on firm ground that reduced MHD accurately captures the field-perpendicular dynamics of the strong turbulent cascade. 

\subsection{Parameter Regime}

We aim to simulate strongly nonlinear, field-guided MHD turbulence in which $v_{rms}\approx b_{rms} \ll B_0$. If we choose the amplitude of the driving so that $v_{rms} \approx 1$, our previous work has shown that the universal properties then set in when $B_0 \gtrsim 3$ (e.g.~Ref.~\onlinecite{mason_cb06}). In all of the simulations reported in this paper we have set $B_0=5$.

Although weak turbulence eventually becomes strong as the cascade proceeds to smaller scales, failing to establish the critical balance condition at the driving scales lengthens the transition region and therefore shortens the inertial range.  An efficient way of driving strong turbulence while maintaining an inertial range that is as extended as possible is to elongate the box in the direction of the guide field and to drive the lowest field-parallel and field-perpendicular wavenumbers $k_{\|}=2\pi/L_{\|}$, $k_{\perp}=2\pi/L_{\perp}$. Equation~(\ref{eq:critical_balance}) is then satisfied at the forcing scales provided that 
\begin{equation}
L_{\perp}/{L_\|} \sim z^\mp_{\perp}/B_0
\end{equation}
Here we fix $L_{\perp}=2\pi$. In the balanced case in which the turbulence is driven so that $z^+ \sim z^- \sim 1$, the above condition can then be satisfied by taking $L_{\|} \sim B_0L_\perp$. In fact, the RMHD description already assumes that $B_0 \gg b_{rms}$ and therefore all RMHD simulations with $L_\|=B_0L_{\perp}$ produce equivalent results.

The random forcing mechanisms constitute independent driving of both Els\"asser populations, which allows us to vary the $z^\pm$ fluxes independently and study both the balanced and imbalanced regimes. As has been well documented in the literature, the particular mechanism of large scale driving is not essential for the scaling of the inertial interval (see, e.g., Refs. \onlinecite{muller_g05, mason_cb08}).  Our driving mechanism for the MHD case has no component along $z$ and is solenoidal in the $x-y$ plane. 
The random values of the Fourier coefficients of the forces inside the range of wavenumbers $1 \le |k_x|,|k_y| \le 2$ and $(2\pi/L_{\|}) \le |k_z| \le n_\|(2\pi/L_\|)$ with $n_\|=1$ or $2$ are refreshed independently on average every $0.1 L_{\perp}/(2\pi v_{rms})$ time units (i.e. the force is updated approximately $10$ times per large-scale turnover time) with the amplitude of the force being chosen so that the resulting rms velocity fluctuations are of order unity. The variances $\sigma^2_\pm=\langle|\vec{f^\pm}|^2\rangle$ control the average rates of injection into the $z^+$ and $z^-$ fields. We take $\sigma_+ \ge \sigma_-$ and in the statistically state we measure the degree of imbalance through the parameter $\sigma_c=H^c/E=(E^+-E^-)/(E^++E^-)$. Thus balanced turbulence corresponds to $\sigma_c=0$, while the maximally aligned/imbalanced case corresponds to $\sigma_c=1$.

The numerical resolution and Reynolds number are intricately related and one would like to conduct simulations with as high a resolution and $Re$ as possible. The upper bound on the resolution is determined by the availability of computing resources, with a doubling of the number of mesh points in each of the three directions resulting in an increase in the computational cost by a factor of 16. For a chosen calculation size, the optimal Reynolds number $Re=v_{rms}l_0/\nu \approx 1/\nu$ can be established via a convergence study. Once a satisfactory $Re$ has been established at low resolution, an estimate of the permitted value for higher resolution studies can be obtained from the scaling for the number of mesh points $N \sim Re^{\beta}$ where $\beta=2/3$ for the spectral exponent $p=-3/2$ ($\beta=3/4$ for $p=-5/3$).  A further note is required concerning the field-parallel resolution: Since the turbulent cascade is anisotropic we also allow for the numerical grid to be anisotropic, i.e. the spatial discretisation is performed on a grid with a resolution of $N_{\perp}^2 \times N_{\|}$ mesh points. For cases in which $N_{\|}$ is much less than $N_{\perp}$ we replace the diffusion operator in equations~(\ref{eq:mhd-elsasser}) and (\ref{eq:rmhd-elsasser}) with $\nu(\partial_{xx}+\partial_{yy})+\nu_{\|}\partial_{zz}$.  

We note that hyperdiffusion is sometimes used to artificially extend the inertial range, i.e. the diffusive terms in equations (\ref{eq:mhd-elsasser}) and (\ref{eq:rmhd-elsasser}) are sometimes replaced with the higher-order operator $(-1)^{n-1}\nu_n \nabla^{2n}$ for $n>1$. In hydrodynamic turbulence, hyperdiffusion is known to lead to a bottleneck effect (or a pile up of energy at large wavenumbers) that can ultimately affect the inference of scaling laws within the inertial range.  Its effects on the inertial range dynamics for MHD are not well understood, with simulations by different groups producing different results \cite{beresnyak_l10,chen_etal11b}. Since our aim is to conduct as clean a simulation as possible, and since our task of differentiating between two very similar scaling exponents is difficult enough as it is, we prefer not to incorporate hyperdiffusion into our simulations.

Finally, it is important to have a large enough statistical ensemble from which averages will be computed. All of the results reported in this paper constitute averages over tens of snapshots of the system (typically 50-100 snapshots), with each snapshot being separated by an interval of the order of the eddy turnover time. 

\subsection{Numerical Method}

We solve equations (\ref{eq:mhd-elsasser},\ref{eq:mhd-div}) and (\ref{eq:rmhd-elsasser},\ref{eq:rmhd-div}) on a triply periodic domain using standard pseudospectral methods. The time-advancement of the diffusive terms is carried out exactly using the integrating factor method, while the remaining terms are treated using a third-order Runge-Kutta scheme. For a detailed description of the numerical method, see, e.g.,~Ref.~\onlinecite{cattaneo_ew03}. We have conducted a number of MHD and RMHD simulations in both the balanced and imbalanced regimes. The parameters for each of the simulations are shown in Table~\ref{tab:sim_list}. 

\begin{table}
\begin{ruledtabular}
\caption{\label{tab:sim_list} The parameters for each of the simulations. The first letter in the case number denotes the regime: M (MHD; eq.~(\ref{eq:mhd-elsasser})) or R (RMHD; eq.~(\ref{eq:rmhd-elsasser})). In all cases $\nu_\|=\nu$, except for cases R2 and R3 where $\nu_\|=2.5\nu$. }
\begin{tabular}{lcccccc}
Case & $N_{\perp}$ & $N_{\|}$ & $L_{\|}/L_{\perp}$ & $n_\|$ & $Re=1/\nu$ & $\sigma_c$\\
\hline
M1 & 512 & 512 & 6  & 1  & 1800  & 0 \\
R1 & 512 & 512 & 6  & 1  &  1800  & 0 \\
M2 & 1024 & 256 & 10  & 1  & 5600 & 0.5\\
R2 & 1024 & 256 & 10  & 1  & 5600  & 0.5 \\
M3 & 256 & 256 & 6  & 1  & 800 &  0\\
M4 & 1024 & 1024 & 6  & 1  & 3200  & 0\\
R3 & 2048 & 512 & 10  & 1  & 14000 & 0.5\\
R4 & 512 & 256 & 10  & 1  & 2200 &  0.5 \\
M5 & 512 & 512 & 5 & 1  &  1800  & 0\\
M6 & 512 & 512 & 5 & 1  & 3200  & 0\\
M7 &  512 & 512 &  5  & 1  & 4000  & 0 \\
M8 & 1024 & 1024 & 5  & 1  & 4000  & 0 
\label{tab:sim_list}
\end{tabular}
\end{ruledtabular}
\end{table}

\section{\label{sec:Results} Results}

\subsection{Comparing RMHD with MHD}
\label{sec:compare}

\begin{figure*} [tbp]
\begin{center}
\resizebox{0.45\textwidth}{!}{\includegraphics{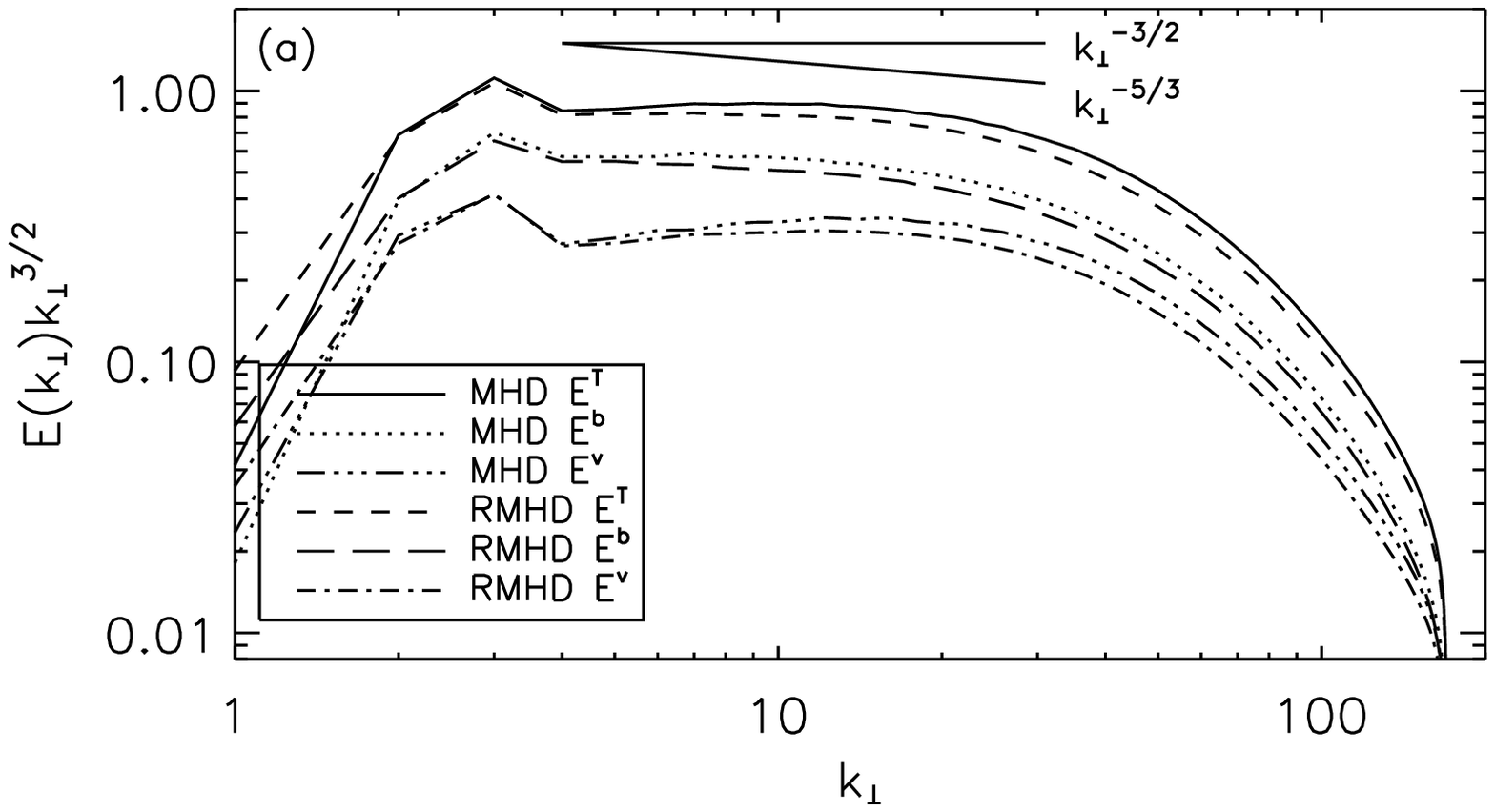}}
\resizebox{0.45\textwidth}{!}{\includegraphics{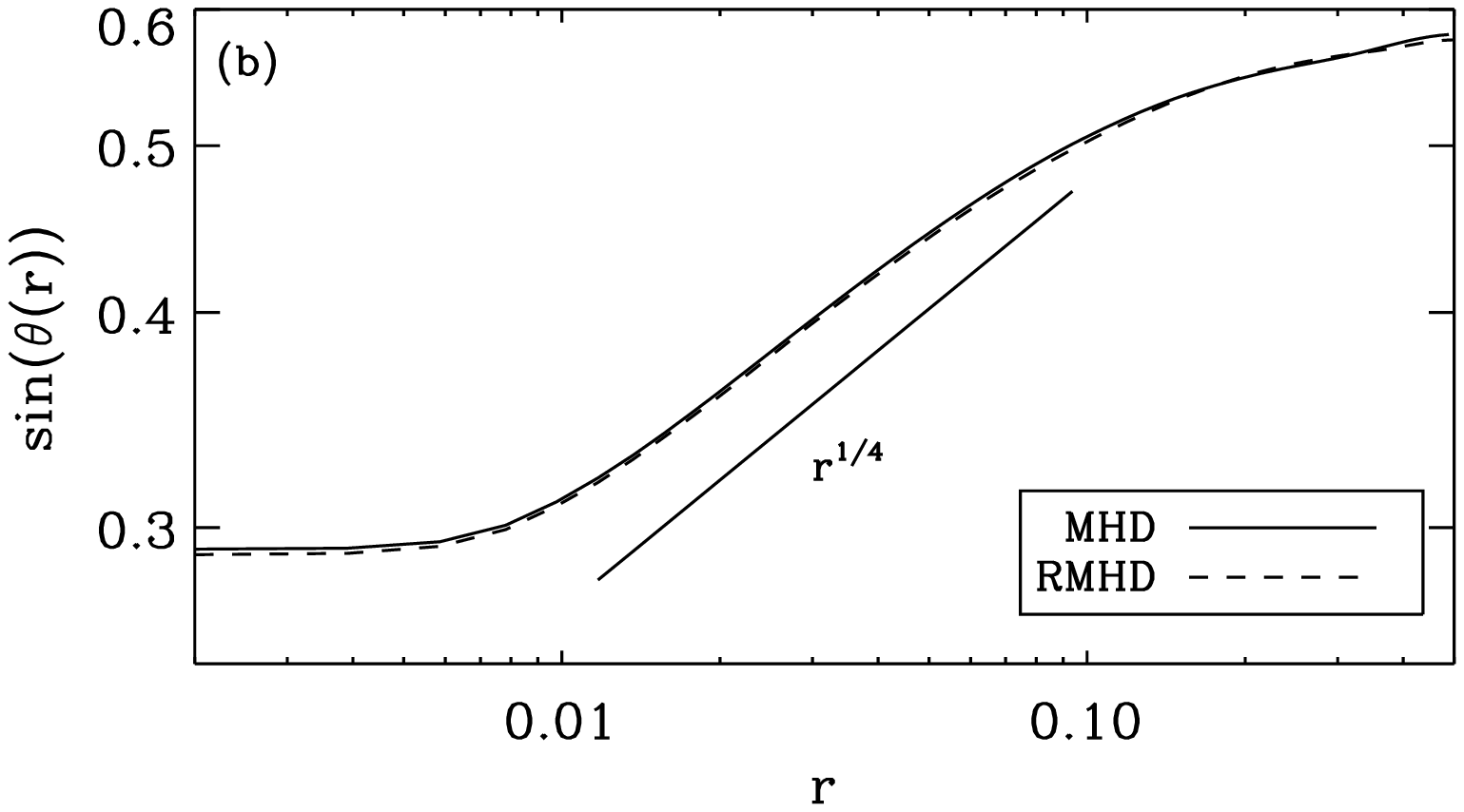}}
\end{center}
\caption{A comparison of the MHD and RMHD field-perpendicular energy spectra (a) and the alignment angle (b) for balanced turbulence (Cases M1 and R1).}
\label{fig:MHD_RMHD_balanced}
\end{figure*}

Here we directly compare the numerical solution of the full MHD system (\ref{eq:mhd-elsasser},\ref{eq:mhd-div}) with the RMHD system (\ref{eq:rmhd-elsasser},\ref{eq:rmhd-div}) in both the balanced and imbalanced regimes. For the balanced case we compare simulations M1 and R1. For the imbalanced case we compare the results of M2 and R2, for which the ratio of the Els\"asser energies $E^+/E^-=(1+\sigma_c)/(1-\sigma_c)\approx 3$. 

Shown in Figure~\ref{fig:MHD_RMHD_balanced}a are the 
field-perpendicular energy spectra for the velocity $E^v(k_{\perp})$, the magnetic field $E^b(k_{\perp})$ and the total spectrum $E^T=E^v+E^b$ for the case of balanced turbulence. Here
\begin{equation}
E^q(k_{\perp})=\frac{1}{2}\langle |\vec{\hat q}(k_{\perp})|^2\rangle k_{\perp},
\end{equation}
where $q$ represents either the velocity or the magnetic field, $\vec{\hat q}(k_{\perp})$ is the two-dimensional Fourier transformation of $\vec{q(\vec{x})}$ in a plane perpendicular to $\vec{B_0}$ and $k_{\perp}=(k_x^2+k_y^2)^{1/2}$. The average is taken over all field-perpendicular planes in the data cube and then over all data cubes (i.e.~snapshots). The resulting field-perpendicular spectrum is equivalent to that obtained by integrating the three-dimensional Fourier spectrum over $k_z$. It is clear that the three MHD and RMHD spectra are very similar beyond the forcing scales (the difference at $k_{\perp}=1,2$ is due to a slight difference in the forcing mechanisms in the MHD and RMHD cases, with the RMHD system (\ref{eq:rmhd-elsasser}) actually being solved for the Els\"asser potentials $\phi^\pm$, where $\vec{z}_{\perp}^\pm=\vec{\hat e_z}\times \nabla \phi^{\pm}$, and hence incompressibility being satisfied automatically, while for MHD the forces are constrained to be solenoidal). In particular, in both cases the total energy spectra are more closely matched with $E^T(k_{\perp}) \propto k_{\perp}^{-3/2}$ than $k_{\perp}^{-5/3}$, with the inertial range corresponding to $4 \lesssim k_{\perp} \lesssim 20$. It is also the case that for both MHD and RMHD the magnetic spectrum is slightly steeper than $k_{\perp}^{-3/2}$ and the velocity spectrum is slightly flatter. This interesting finding is pursued in more detail in Ref.~\onlinecite{boldyrev_pbp12}. 

Figure~\ref{fig:MHD_RMHD_balanced}b compares the alignment angle $\theta(r)$ 
between the shear Alfv\'en velocity and magnetic fluctuations, which we define through the ratio
\begin{equation}
\theta_r \approx \sin(\theta_r)=\frac{\langle \delta \vec{ \tilde v_r} \times \delta \vec{\tilde b_r} \rangle} {\langle |\delta \vec{\tilde v_r}| |\delta \vec{\tilde b_r}|\rangle}.
\end{equation}
Here $\vec{\delta v_r}=\vec{v(x)}-\vec{v(x+r)}$ and $\vec{r}$ is a point separation vector in the plane perpendicular to the background magnetic field $\vec{B_0}$. The average is taken over all points in a field-perpendicular slice, over all such slices in the data cube, and over all data cubes. In the MHD case the pseudo Alfv\'en fluctuations are eliminated by removing the parts of $\delta \vec{v_r}$ and $\delta \vec{b_r}$ that are in the direction of the local guide field, i.e. we construct $\delta \vec{ \tilde v_r}=\delta \vec{ v_r}-(\delta \vec{v_r} \cdot \vec{n}) \vec{n}$ where $\vec{n}=\vec{B(x)}/|\vec{B(x)}|$. For RMHD the projection is not necessary. Figure~\ref{fig:MHD_RMHD_balanced}b illustrates that the alignment angle in the MHD and RMHD regimes are almost indistinguishable and that both are in excellent agreement with the theoretical prediction $\theta \propto r^{1/4}$.

\begin{figure} [tbp]
\resizebox{0.5\textwidth}{!}{\includegraphics{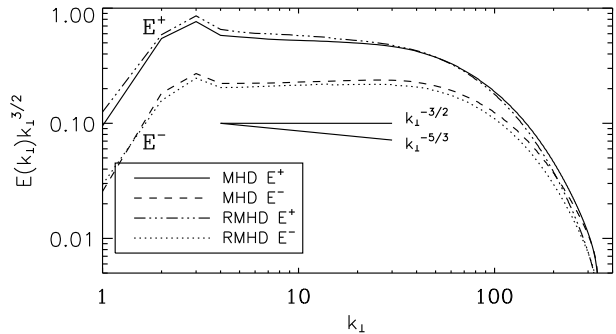}}
\caption{A comparison of the field-perpendicular energy spectra $E^\pm$ for the case of imbalanced MHD and RMHD turbulence (Cases M2 and R2).}
\label{fig:MHD_RMHD_imbalanced_spec}
\end{figure}

A comparison of the field-perpendicular Els\"asser spectra $E^\pm(k_{\perp})$ in the MHD and RMHD regimes of imbalanced turbulence is shown in Figure~\ref{fig:MHD_RMHD_imbalanced_spec}. Again the agreement between the two systems is very good. In particular, we find that weaker field obeys $E^-(k_{\perp}) \propto k_{\perp}^{-3/2}$, while $E^+(k_{\perp})$ is slightly steeper at this resolution. In the next section we illustrate that the spectrum for the stronger field becomes flatter as the Reynolds number increases, implying that the universal regime has not yet been reached. Nonetheless, it is evident that the MHD and RMHD systems behave similarly (see also Ref.~\onlinecite{perez_etal12}). 

In summary, we find that in both the balanced and imbalanced regimes, the pseudo Alfv\'en waves do not significantly impact the strong turbulent cascade and RMHD accurately captures the MHD dynamics.
 
\subsection{Increasing the numerical resolution and Re}

\begin{figure*} [tbp]
\resizebox{0.45\textwidth}{!}{\includegraphics{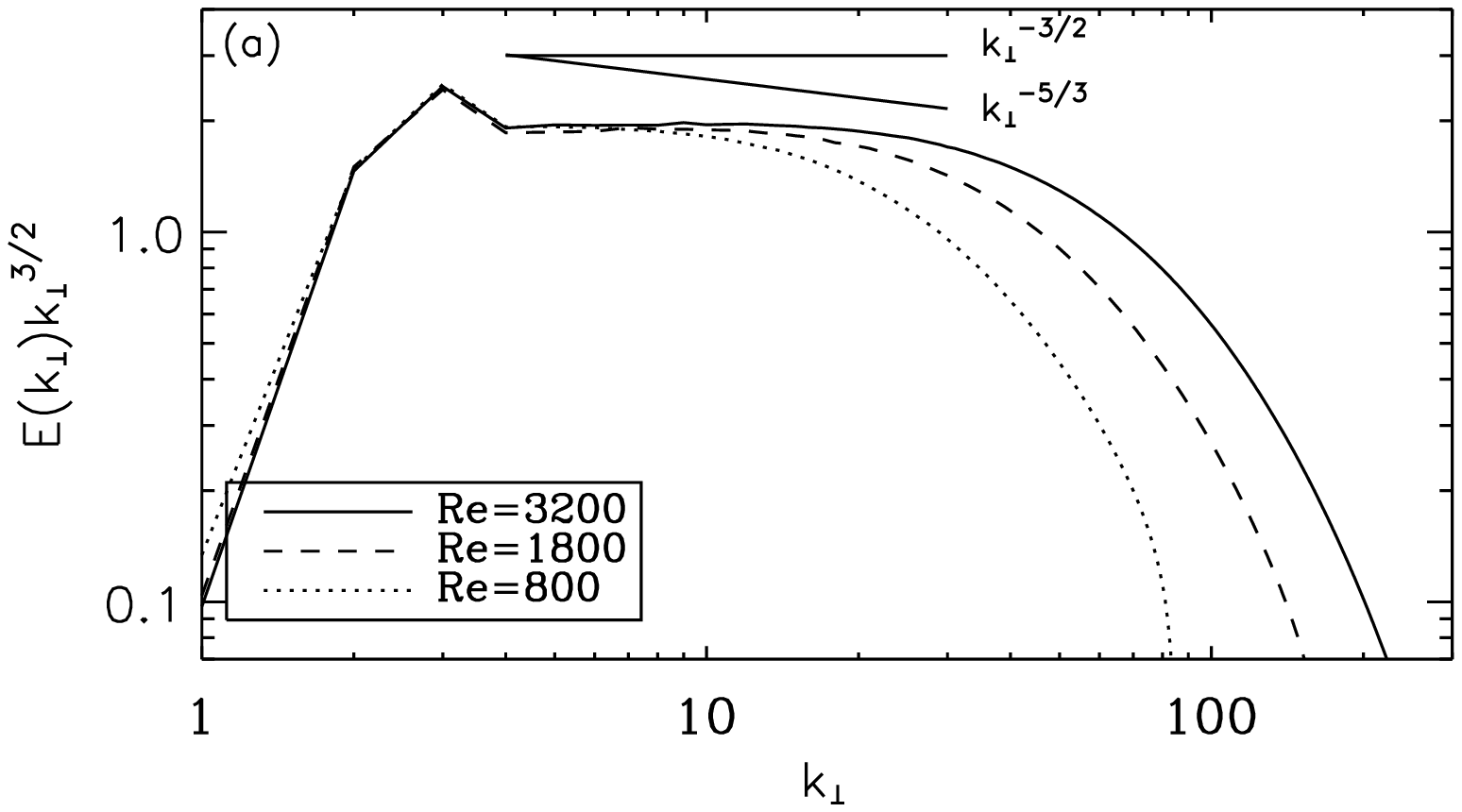}}
\resizebox{0.45\textwidth}{!}{\includegraphics{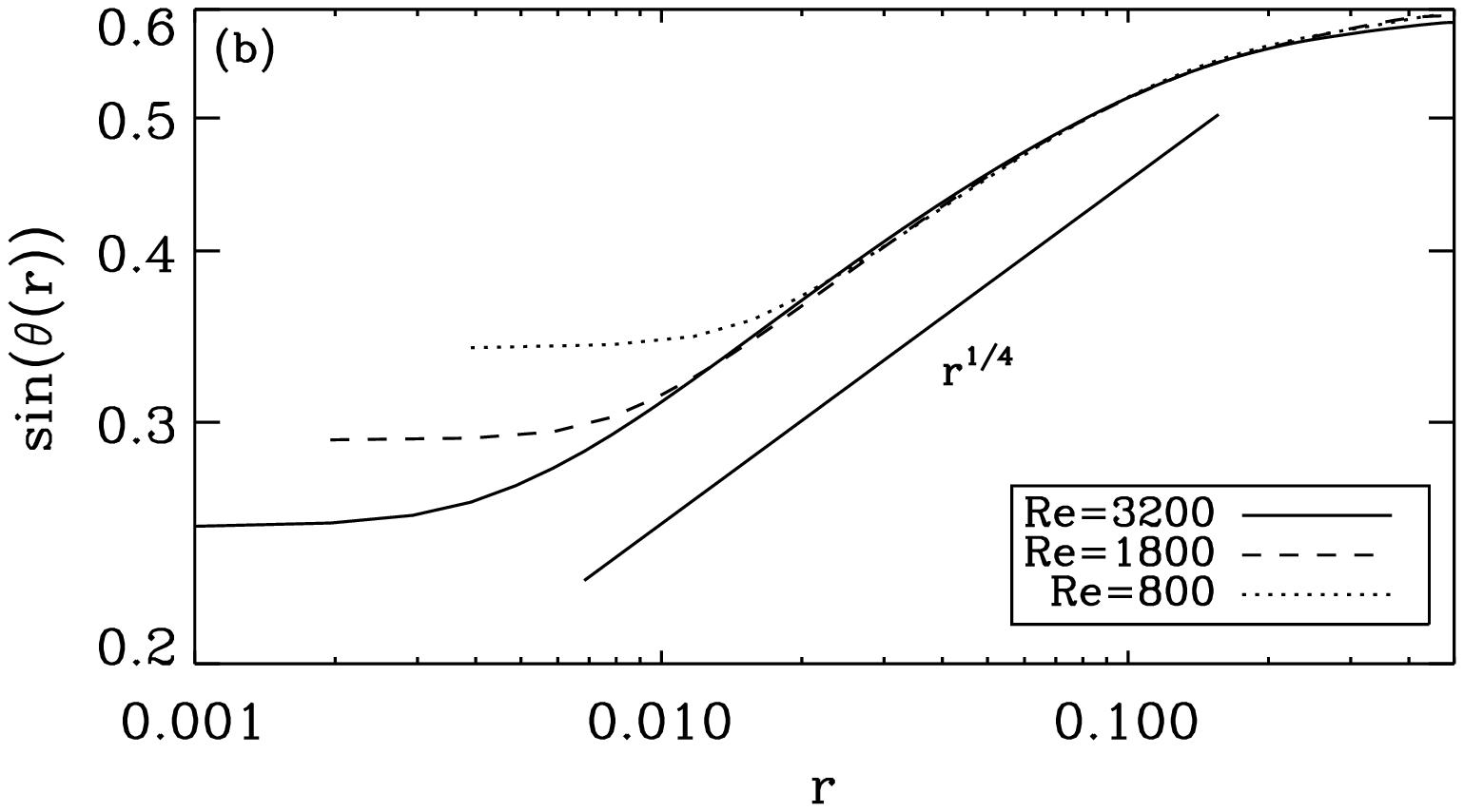}}
\caption{The total field-perpendicular energy spectrum (a) and the alignment angle (b) for a series of MHD calculations corresponding to increasing the numerical resolution and the Reynolds number (Cases M3, M1 and M4).}
\label{fig:spectra_res}
\end{figure*}

We now investigate the robustness of the scalings laws as the numerical resolution and the Reynolds number increase and hence the extent of the inertial range grows.

Illustrated in Figure~\ref{fig:spectra_res} are the results for a set of balanced MHD simulations corresponding to a four-fold increase in the numerical resolution. We note that all of the simulations in this set have a strong guide field, the computational domain is elongated in the $z$-direction in proportion to $B_0$, the number of mesh points in the field-parallel and field-perpendicular directions are equal, and the turbulence is excited in the strong state (i.e. the wavenumbers driven and the forcing correlation time are chosen so that the critical balance condition holds at the driving scales). Figure~\ref{fig:spectra_res}a shows that the total field-perpendicular energy spectrum maintains the scaling $E^T(k_{\perp}) \propto k_{\perp}^{-3/2}$ as the inertial range grows. We also note that the spectra fall off smoothly at large wavenumbers, i.e. that the `bottleneck effect' that leads to a pile up of energy at small scales is not present. Figure~\ref{fig:spectra_res}b illustrates that the scaling of the alignment angle is also in excellent agreement with the theoretical prediction $\theta_r \propto r^{1/4}$, with the extent of the region over which this scaling holds increasing as the resolution (and $Re$) increases. In fact, as explained in Ref.~\onlinecite{mason_etal11}, we believe that the alignment angle displays a significantly extended self-similar region that persists deep into the dissipation range, with the saturation of the $r^{1/4}$ scaling being controlled by the smallest resolved scale of the simulation and hence decreasing by a factor of 2 as the resolution doubles.

\begin{figure} [tbp]
\resizebox{0.5\textwidth}{!}{\includegraphics{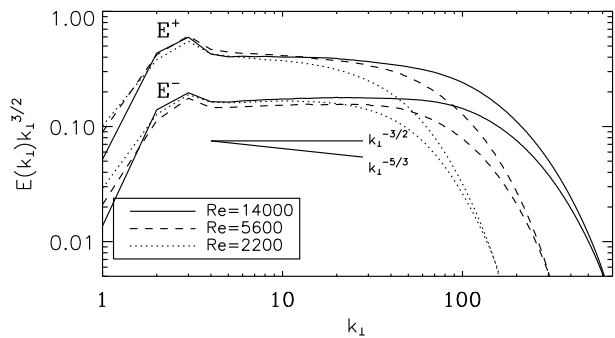}}
\caption{The field-perpendicular Els\"asser spectra for a series of RMHD calculations corresponding to increasing the numerical resolution and the Reynolds number (Cases R4, R2 and R3).}
\label{fig:imbalanced_spectra_res}
\end{figure}

The effects on the Els\"asser spectra $E^\pm(k_{\perp})$ of increasing the numerical resolution by a factor of four in the case of imbalanced RMHD turbulence are shown in Figure~\ref{fig:imbalanced_spectra_res}. It is seen that $E^-(k_{\perp}) \propto k_{\perp}^{-3/2}$ for all Reynolds numbers considered. However, the scaling of the stronger Els\"asser field is difficult to pin down at these resolutions. Indeed, $E^+(k_{\perp})$ appears to flatten as $Re$ increases, and thus the universal regime has not yet been reached. Indeed, this is consistent with the Reynolds number for the stronger Els\"asser field $Re^+ \sim z^-l/\nu$ being smaller than that for the weaker field. However, since the spectra are pinned\cite{grappin_pl83, lithwick_g03} at the dissipation scale and anchored at the driving scale we proposed in Ref.~\onlinecite{perez_etal12} that at sufficiently large Reynolds numbers the two spectra will become parallel and obey the scaling $E^\pm(k_{\perp}) \propto k_{\perp}^{-3/2}$. 

\subsection{Contamination of the scaling laws}

Having established above that for balanced turbulence the inertial range scalings are robust in optimally designed, carefully conducted numerical simulations, we now proceed to describe how numerical measurements of both the spectral exponent and the alignment angle can be contaminated. We emphasize that the scaling properties are spoilt entirely due to numerical rather than physical effects. The numerical errors can originate either at small scales and then back scatter to ultimately contaminate the inertial range, or they can arise at large scales due to less than optimal numerical settings for studying strong magnetised turbulence. 

For strong balanced turbulence, it is known that numerical measurements of the total field-perpendicular energy spectrum can yield an exponent steeper than $-3/2$ when the properties of the driving are poorly controlled. For example, it was shown in Ref.~\onlinecite{perez_b08} that, in the RMHD regime, as the number of modes driven in the field-parallel direction is increased the spectrum becomes steeper, eventually approaching the weak turbulence result $E (k_{\perp}) \sim k_{\perp}^{-2}$ (Refs.~\onlinecite{ng_b96,galtier_nnp00}). A similar effect was found in Ref.~\onlinecite{mason_cb08} for MHD turbulence. Such effects are a result of the limited extent of the inertial range. Since present day computing power limits the inertial scales to approximately a decade in length, being able to reach the universal regime strongly relies on being able to limit the extent of the transition from the driving scales as much as possible. Such a contamination of the inertial scales would not arise if there was a significant separation between the driving scales and the small scale turbulence, as is the case in astrophysical settings where the Reynolds numbers are estimated to be orders of magnitude greater than the few thousands that present day computing power permits. 

\begin{figure}
\resizebox{0.5\textwidth}{!}{\includegraphics{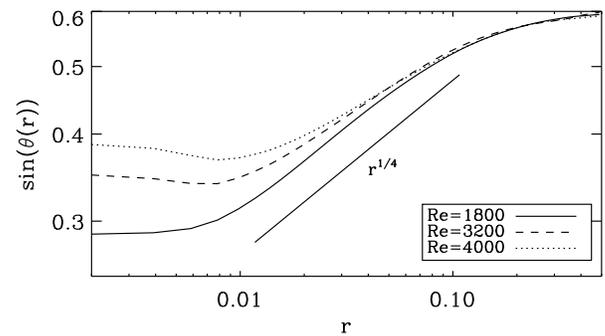}}
\caption{\label{fig:angle_Re} The alignment angle as the Reynolds number increases at fixed resolution (Cases M5, M6 and M7).}
\end{figure}

\begin{figure} [tbp]
\resizebox{0.5\textwidth}{!}{\includegraphics{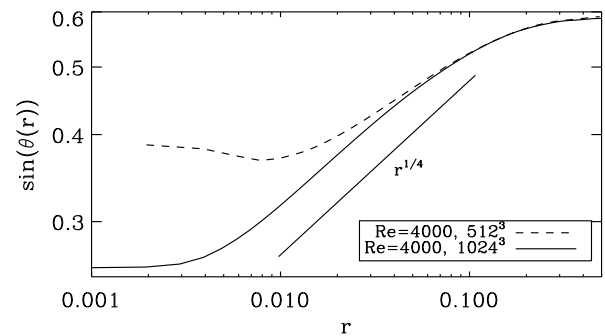}}
\caption{The scaling of the alignment angle at high $Re$ can be recovered by increasing the numerical resolution (Cases M7, M8).}
\label{fig:compare_angle_Re4000}
\end{figure}

In Ref.~\onlinecite{mason_etal11}, we explained how the alignment angle in balanced MHD turbulence can be spoilt by a strongly decreased field parallel resolution. When the simulation is well resolved, alignment persists to much smaller scales than those over which the energy spectra display a power law behaviour. In Ref.~\onlinecite{mason_etal11} we proposed that this could be due alignment being measured as the ratio of two structure functions whose non-universal features cancel. Consequently, accurate measurement of the extended scaling behaviour relies on adequately resolving the small scale physics. Clearly this is spoilt when the resolution degrades. Figure~\ref{fig:angle_Re} illustrates that a similar contamination occurs when the Reynolds number is pushed to the extreme.  However, it is important that such behaviour should not be interpreted as a breakdown of the alignment mechanism at high $Re$. Indeed, as is shown in Figure~\ref{fig:compare_angle_Re4000}, excellent agreement with the $r^{1/4}$ scaling can be recovered by increasing the numerical resolution. A similar flattening of the alignment angle was shown in Ref.~\onlinecite{beresnyak_11} for a simulation with a reduced field-parallel resolution and sixth-order hyperviscosity. 


\section{\label{sec:Discussion} Discussion}

High-resolution direct numerical simulations play a key role in developing the theory for strong field-guided MHD turbulence. Success relies on working hard to harness all of the available computational power into reaching the universal regime. When this is achieved, the numerical data can be used to identify the preferred physical description when there exist conflicting theoretical models. Furthermore, the simulations often reveal interesting new effects that subsequently guide further theoretical progress. 

We have discussed how the numerical setup can be optimised in order to accurately simulate strong MHD turbulence. By direct comparison with the full MHD system, we have also verified that the reduced MHD equations accurately model the turbulent cascade while being twice as computationally efficient. We have discussed how balanced turbulence comprises local domains of highly aligned and anti-aligned velocity and magnetic fluctuations, obeying the theoretically predicted alignment scaling $\theta \propto r^{1/4}$. Consistent with the theory of dynamic alignment is the finding that the total field-perpendicular energy spectrum $E(k_{\perp}) \propto k_{\perp}^{-3/2}$, with this scaling being maintained throughout a four-fold increase in numerical resolution. A similar set of simulations of imbalanced turbulence at steadily increasing Reynolds numbers up to $Re=14000$ have shown that the weaker Els\"asser field obeys the same scaling, $E^-(k_{\perp}) \propto k_{\perp}^{-3/2}$. While a convincing result for the stronger Els\"asser field must await higher resolution tests, we propose that in the high $Re$ limit the two spectra will become parallel and attain the scaling $E^\pm(k_{\perp}) \propto k_{\perp}^{-3/2}$.

We have also shown how sensitive numerical measurements can be to the design of the simulation. 
In particular, we have shown how poorly resolved simulations will spoil measurements of the alignment angle and we have discussed how the spectral exponent can be sensitive to the physics of the large scale driving. The problems are due to the fact that computational power severely limits the extent of the inertial range, and as a consequence numerical effects can hamper our ability to infer the correct physics. In the ideal case, any contamination at the driving or dissipative scales should not be felt sufficiently deep in the inertial range. Until enormous increases in computational power permit the investigation of such a regime, one must strive to ensure that the simulation is as clean as possible and that the numerical results are well converged. 

\begin{acknowledgments}
This research was supported by the NSF sponsored Center for Magnetic Self-Organization in Laboratory and Astrophysical Plasmas at the University of Chicago and the University of Wisconsin - Madison, the US DoE awards DE-FG02-07ER54932, DE-SC0003888, DE-SC0001794, and the NSF grants PHY- 0903872 and AGS-1003451. 
The simulations were conducted and analysed using `Intrepid' at Argonne Leadership Computing Facility at Argonne National Laboratory, `Kraken' at the National Institute for Computational Sciences, and `Beagle' and `PADS' at the Computation Institute at the University of Chicago.
\end{acknowledgments}

\end{document}